\documentclass[%
 reprint, amsmath,amssymb,
 aps,superscriptaddress,showkeys]{revtex4-2}
\usepackage{graphicx}
\usepackage{dcolumn}
\usepackage{bm}
\usepackage{xcolor}
\usepackage{mathtools}
\usepackage[
citecolor=blue,
colorlinks,
linkcolor=blue,
urlcolor=blue,
]{hyperref}
\usepackage[normalem]{ulem}

\begin{document}

\title{Periodic orbits in \textcolor{black}{deterministic discrete-time} evolutionary game dynamics: An information-theoretic perspective}

\author{Sayak Bhattacharjee}
\email{sayakb@iitk.ac.in}
\affiliation{%
Department of Physics, Indian Institute of Technology Kanpur, Uttar Pradesh 208016, India.
}%
\author{Vikash Kumar Dubey}%
\email{vdubey@iitk.ac.in}
\affiliation{%
Department of Physics, Indian Institute of Technology Kanpur, Uttar Pradesh 208016, India.
}%
\author{Archan Mukhopadhyay}%
\email{ a.mukhopadhyay@bham.ac.uk}
\affiliation{%
School of Computer Science, University of Birmingham, Birmingham B15 2TT, United Kingdom}%
\author{Sagar Chakraborty}
\email{sagarc@iitk.ac.in}
\affiliation{%
Department of Physics, Indian Institute of Technology Kanpur, Uttar Pradesh 208016, India.
}%

\date{\today}

\begin{abstract}
Even though existence of non-convergent evolution of the states of populations in ecological and evolutionary contexts is an undeniable fact, insightful game-theoretic interpretations of such outcomes are scarce in the literature of evolutionary game theory. \textcolor{black}{As a proof-of-concept}, we tap into the information-theoretic concept of relative entropy in order to construct a game-theoretic interpretation for periodic orbits in a wide class of \textcolor{black}{deterministic discrete-time} evolutionary game dynamics, \textcolor{black}{primarily investigating the two-player two-strategy case}. Effectively, we present a consistent generalization of the evolutionarily stable strategy---the cornerstone of the evolutionary game theory---and aptly term the generalized concept: information stable orbit. The information stable orbit captures the essence of the evolutionarily stable strategy in that it compares the total payoff obtained against an evolving mutant with the total payoff that the mutant gets while playing against itself. Furthermore, we discuss the connection of the information stable orbit with the dynamical stability of the corresponding periodic orbit.
\end{abstract}
\keywords{Evolutionary games, evolutionarily stable strategy, replicator map, escort-incentive dynamics, periodic orbits, relative entropy.}
\maketitle

\section{\label{intro} Introduction}
Many ecological systems are known to exhibit cyclic evolution of the abundance of the constituent species---well-studied examples include the snowshoe-hare--lynx system~\cite{harelynx}, wolf-moose system~\cite{wolf-moose} and lemming populations~\cite{Fauteux2015}. The simplest explanation for such a cyclic behaviour is mostly attributed to predation, which may be modelled by the corresponding Lotka--Volterra equation~\cite{Lotka1920}. Since the Lotka--Volterra equation is mathematically mappable~\cite{Hofbauer1998} onto the replicator equation~\cite{taylorjonker, Schuster1983}---a paradigmatic evolutionary dynamic in the theory of evolutionary games~\cite{Smith1982}---the relevance of oscillatory dynamics in evolutionary game theory is worth pondering. 

\textcolor{black}{One of the central ideas} in evolutionary game theory is that of evolutionarily stable strategy (ESS)~\cite{SMITH1973}, which happens to be Nash equilibrium~\cite{Nash1950, Nash1951} as well. Through the folk theorem (and other similar theorems)~\cite{2014_CT_PNAS} of evolutionary game theory, the convergent outcomes---represented by a fixed point in phase space---of the replicator dynamics can be interpreted as Nash equilibrium and ESS whenever achievable. These game-theoretic interpretations of the fixed points can be further supplemented with interesting information-theoretic connection: A fixed point that is an ESS is locally asymptotically stable and  an appropriately constructed Kullback--Leibler (KL) divergence (also called relative entropy)~\cite{Cover2005} is a Lyapunov function~\cite{jordan-smith} for the fixed point, i.e., the nonnegative function decreases with the passage of time~\cite{baez_entropy} until the dynamics converges onto the fixed point eventually.

Nevertheless, there is a dearth of literature about the game-theoretic and the information-theoretic interpretations for non-convergent oscillatory outcomes. In this paper, we contribute to this area. For pragmatic reasons, it is rather convenient to work with discrete-time replicator equation (replicator map) as even with a population with two strategies (phenotypes), one can witness cyclic behaviour \cite{vilone,pandit}. In fact, using the replicator map, it was shown earlier~\cite{archan_periodic} that a game-theoretic interpretation of periodic orbits is possible---an extension of ESS, called heterogeneity stable orbit (HSO), was proposed. However, unlike the ESS, the concept of the HSO does not consider fitness as the quantity to be optimized during natural selection; rather it concerns itself with the optimization of a weighted fitness---the weight being heterogeneity. The heterogeneity is the probability that two arbitrarily chosen members of the population belong to two different phenotypes. Although a justification for the HSO could be found in its connection with the stable periodic orbits just as ESS is connected with stable fixed points, going beyond the well-accepted notion of fitness to a weighted fitness appears unconventional. Moreover, it is not clear if HSO conforms with the principle of decreasing relative entropy~\cite{Cover2005}.

\textcolor{black}{A natural question emerges at this point: Using the} notion of decreasing relative entropy during the evolution of the system, \textcolor{black}{does there exist} a natural extension of ESS for the case of periodic orbits such that fitness (sans any weight) is the object of optimization? \textcolor{black}{We answer this question affirmatively in this paper by the introduction of the extension of ESS termed appropriately as Information Stable Orbit (ISO)}.  We also find an encouraging connection of this game-theoretic concept---obtained though information-theoretic ideas---with the dynamical system theoretic concept of stability of  periodic orbits. Our investigation is not restricted to only the replicator map, rather we expand our investigations to encompass a much wider class of evolutionary dynamics like incentive~\cite{fryer2013existence}  and escort-incentive dynamics~\cite{harper_timescales} (which account for many well-studied dynamics like logit~\cite{Wagner2013} and best-reply~\cite{2003_Cressman}).

\textcolor{black}{Given the rather technical nature of the results in our work, below we first provide a succinct outline of the key aspects of our work and the main results.}

\subsection{Outline of the paper}
\textcolor{black}{As noted in the introduction, the central result of this work is a generalization of the game-theoretic notion of evolutionarily stable strategies (ESS)---traditionally only defined for fixed points---to periodic outcomes in discrete-time evolutionary dynamics. To do this, we revisit the relative entropy minimization principle for evolutionary dynamics~\cite{baez_entropy,bomze1991cross, karev2010replicator}. As per this principle, if an evolving population state is close to an equilibrium point, the relative entropy (quantified using the Kullback-Leibler (KL) divergence, defined in Eq.~\eqref{eq:kldiv}) monotonically decreases. This heuristic principle may be considered as an application of Kullback's general proposal of the principle of minimum discrimination information---which has been used successfully in various contexts~\cite{olivares2007quantum, qian1991relative, floerchinger2020thermodynamics}---to evolutionary systems. In fact, this may be interpreted quite physically: Darwinian evolution ensures that the {amount of biological information} produced during evolution, as compared to the equilibrium state, is optimized~\cite{karev2010replicator}. }

\textcolor{black}{In Sec.~\ref{fixed point}, we explicitly demonstrate how minimization of the KL-divergence yields the well-known game-theoretic stability criteria (such as ESS) for fixed points in evolutionary dynamics. While such results have been explored recently for continuous-time dynamics, we extend this to discrete-time dynamics. In addition to the paradigmatic replicator map (see Eq.~\eqref{eq:ret1}), we also discuss two generalized classes of evolutionary dynamics as well---the incentive and escort-incentive dynamics. The incentive dynamic (Eq.~\eqref{incentive_map}) accommodates for various updation protocols of incentive in evolutionary games, while the escort-incentive dynamic  (Eq.~\eqref{escort-incentive_map}) additionally generalizes the averaging procedure in the incentive map.}  

\textcolor{black}{Our main results are in Sec.~\ref{sec:periodic_orbits}, where we extend this program for periodic orbits in discrete-time evolutionary dynamics. A previous (purely game-theoretic) endeavour~\cite{archan_periodic} to construct a generalization of ESS for periodic orbits in such dynamics was the Heterogeneity Stable Orbit (HSO) (see Eq.~\eqref{eq:hso}), which was also accompanied by a generalization of the Nash equilibrium, known as the Heterogeneity Orbit (HO) (Eq.~\eqref{eq: HE v1}). From a game-theoretic perspective, the hope was straightforward: Just as an ESS state yields the maximum payoff against an arbitrary state in its neighbourhood as compared to the state playing against itself, a generalization of ESS for periodic orbits---which would be a sequence of states---would achieve this in a periodically averaged sense. In other words, we intuit that the periodic orbit would yield the maximum total payoff when played against a sequence of evolving states as opposed to the evolving states playing against themselves. The HSO condition achieves this in a little weak fashion because it involves a weighted payoff, where the weight measures the heterogeneity in the evolving population state. Quite elegantly, our information-theoretic technique, adopted in this paper, achieves our intuition exactly: We do not require a weighted payoff and simply compare the sum of payoffs over a period; thus, generalizing the notion of ESS for periodic orbits in the discrete-time evolutionary dynamics. We dub this generalization the Information Stable Orbit (ISO) and formulate it for the replicator map (Eq.~\eqref{ISO_replicator}), the incentive map (Eq.~\eqref{eq:IISO}) and the escort-incentive map (Eq.\eqref{eq:IEISO}). } 

\textcolor{black}{Finally, before we conclude the work in Section~\ref{sec:discussion}, we analytically indicate how an ISO may correspond to dynamically stable periodic orbit (in Sec.~\ref{sec:dyn_stab}) and verify the analysis for various evolutionary dynamics numerically (in Sec.~\ref{sec:numerical_results}). Having presented the brief outline of the body of this paper, we now discuss the technical details in the body of this paper without further ado.} 

\section{Interpreting Fixed Points}\label{fixed point}
With a view to keeping the presentation of the main ideas of the paper succinct, we confine ourselves to the simplest yet conceptually non-trivial set-up of two-person two-strategy games. Specifically, we consider an infinite unstructured population of two types whose frequencies change over time under some replication-selection rule. Between the individuals (or the players, in game-theoretic terminology), the interactions are assumed random and of two-player kind.  The frequencies of the $i$-th type is denoted by $x_i$ and thus, $\mathbf{x}=\left(x_1,x_2\right)=\left(x_1,1-x_1\right)=(x,1-x)$, in different equivalent notations, is the frequency (or state) vector of the population. \textcolor{black}{Thus, throughout the paper, the states are restricted to the unit interval, i.e., $0\leq x\leq 1$, as discussed briefly in subsection below.} Many kinds of dynamics for the evolution of the frequencies are studied in the literature.

First, let us recall the paradigmatic continuous-time replicator dynamic~\cite{taylorjonker, Schuster1983} in the light of information-theoretic ideas. Of particular use, is the relative entropy or the Kullback--Leibler (KL) divergence given by 
\begin{equation}\label{eq:kldiv}
   D_{\textrm{KL}}(\mathbf{p}||\mathbf{q})\coloneqq\sum_{y}p(y)\ln[{p(y)}/{q(y)}] 
\end{equation}
where $y$ belongs to the support of the probability distributions $\mathbf{p}$ and $\mathbf{q}$. The KL-divergence helps to measure the distance between two probability distributions and herein, we use it as a measure of the distance of an evolving population state from a fixed (equilibrium) population state. \textcolor{black}{Note that the states can be interpreted as distributions, even though the dynamics is not stochastic.} If the fixed population state is chosen as the ESS state, the aforementioned KL-divergence decreases with time~\cite{baez_entropy}. Given that the KL-divergence fits the criteria for being a valid Lyapunov function, one may also interpret this as a statement for dynamical stability. There exists, however, an important nuance to this fact: While the local asymptotic stability and evolutionary stability of a fixed point state are synonymous only for two-strategy games, for games with higher number of strategies, ESS implies local asymptotic stability of the corresponding fixed point but the converse is not necessarily true~\cite{taylorjonker}.  
\subsection{Incentive Dynamics}
In this paper, we are motivated to carry out our investigations with discrete-time dynamics because they are a convenient test-bed for investigating {non-convergent outcomes} like periodic orbits and chaos. {In particular,} {the} following is a version of {the} discrete-time replicator {dynamic}~\cite{1997_BS_JET, 2000_HS_JEE,2000_BV_QJE,2003_Cressman, mcelreath, 2010_M_AEJM, vilone,pandit, archan_periodic, Archan_2020_chaos}, {henceforth referred to as the replicator map}, that exhibits periodic orbits and chaos even in the simple two-strategy case:
\begin{equation}
      \Delta x_i^{(k)}=x_i^{(k)}\left[f_i(\mathbf{x}^{(k)})-\left\langle f(\mathbf{x}^{(k)})\right\rangle \right].\label{eq:ret1}
\end{equation}
Here $\Delta x_i^{(k)}\coloneqq x_i^{(k+1)}-x_i^{(k)}$ represents the difference between consecutive population states and $f$ is the fitness function for the types, with the average population fitness being given by $\left\langle f(\mathbf{x}^{(k)})\right\rangle\coloneqq\sum_{j}x_j^{(k)}f_j(\mathbf{x}^{(k)})$. The superscript denotes the time step. \textcolor{black}{We also note that as per our notation, $f(\mathbf{x})$ denotes a vector with components $f_i(\mathbf{x})$.} For this discrete map, it can be shown that a locally asymptotically stable fixed point implies ESS for two-strategy games~\cite{pandit, pandit_correction}, without its converse being necessarily true. \textcolor{black}{To ensure the solutions of the above discrete replicator map (and all such subsequently introduced maps) are physical (i.e., $0\leq x^{(k)}_i\leq 1$ for all $i$ and $k$), we must restrict the parameter space of the map (here, captured by the function $f$; for details, see Eq.~\eqref{eqn:PayOff_A}) to the strict physical region, as defined in Ref.~\cite{pandit, MCC_JPC_21}.} 

\textcolor{black}{The replicator map in Eq.~\eqref{eq:ret1} phenomenologically models the replication-selection dynamics in line
with the Darwinian tenet of natural selection. We know~\cite{pandit} that its fixed points correspond to the
Nash equilibria \cite{Nash48} and the evolutionarily stable strategies or states \cite{maynard_ess} are associated with the folk theorems of the evolutionary game theory \cite{Cressman10810}, just as in the continuous-time case. However, we remark that this form of the replicator equation should not necessarily be seen just as an Euler discretization of the continuous-time equation: It has its own existence unrelated to its counterpart continuous equation. One can motivate this replicator map from viability selection rules \cite{mcelreath}. Also, this replicator equation appears to model intergenerational cultural transmission \cite{2000_BV_QJE, 2010_M_AEJM}, boundedly
rational players, imitational behaviour in bimatrix cyclic games \cite{2000_HS_JEE}, and reinforcement learning \cite{1997_BS_JET}. Thus, it has applicability even beyond usual evolutionary contexts in the biological systems.}

In fact, a unification of quite a few evolutionary dynamics is mathematically possible through the incentive dynamic~\cite{fryer2013existence,fryer_kullbackleibler, fryer_uniform, harper_stochasticenvironments, harper_timescales, harper_finitepopulations, harper_incentiveprocess, harper_incentiveprocess} whose discrete-time representation is given by 
\begin{equation}\label{incentive_map}
    \Delta x_i^{(k)}=x_i^{(k)}(\delta_{\varphi})^{(k)}_i\equiv\varphi_i(\mathbf{x}^{(k)})-x_i^{(k)}\sum_j\varphi_j(\mathbf{x}^{(k)}),
\end{equation}
where $\bm{\varphi}(\mathbf{x}^{(k)})$ is the incentive vector. The notation $(\delta_{\varphi})^{(k)}_i$ has been introduced for future convenience. \textcolor{black}{The incentive dynamic arises by generalizing the notion of incentive for the players of a game, accounting for different updating procedures utilized by the players.} Choice of different incentive functions, $\varphi$, correspond to different dynamics (such as replicator, projection~\cite{Lahkar2008}, logit~\cite{Wagner2013}, and best response~\cite{Matsui1992,Gilboa1991}) as exhibited in Table~\ref{tab:incentive}. 
\begin{table}
\caption{\label{tab:incentive}Incentive functions for various evolutionary dynamics.}
\begin{ruledtabular}
\begin{tabular}{cc}
Evolutionary Dynamic&Incentive ($\varphi_i(\mathbf{x})$)\\
\hline
Projection & $f_i(\mathbf{x})$\\
Replicator& $x_if_i(\mathbf{x})$\\
Logit & $\frac{\textrm{exp}(\beta f_i(\mathbf{x}))}{\sum_i \textrm{exp}(\beta f_i(\mathbf{x}))}$\footnote{Here, $\beta$ is called the rationality factor, see \cite{Wagner2013}.}\\
Best Reply &$x_i\textrm{BR}_i(\mathbf{x})\footnote{$\textrm{BR}(\mathbf{x})$ is the best response function, a detailed discussion is given in \cite{2003_Cressman}.} $
  \\
\end{tabular}
\end{ruledtabular}
\end{table}
A state of the population is an incentive stable state (ISS)~\cite{harper_timescales}, $\hat{\mathbf{x}}$, if for any $\mathbf{x}$ in the local deleted neighbourhood of $\hat{\mathbf{x}}$, we have 
\begin{equation}
\sum_i \hat{x}_i \frac{\varphi_i(\mathbf{x})}{x_i}>\sum_i\varphi_i(\mathbf{x}).\label{eq:iss}
\end{equation}
ISS reduces to ESS for the replicator equation, given explicitly by $\hat{\mathbf{x}}\cdot f(\mathbf{x})>\mathbf{x}\cdot f(\mathbf{x})$. 

For the two-strategy incentive map---as a generalization of the corresponding result of replicator map~\cite{pandit}---one can quite readily show using linear stability analysis that a locally asymptotically stable fixed point obeys the ISS condition. Moreover, it is an easy exercise to show that just like for the replicator dynamic, the ISS for the continuous incentive dynamic implies that the time derivative of the KL-divergence, $D_{\textrm{KL}}(\hat{\mathbf{x}}||\mathbf{x})$, is negative. In what follows, we observe that negative discrete-time derivative of the KL-divergence, defined as 
\begin{equation}
   \Delta D_{\textrm{KL}}(\hat{\mathbf{x}}||\mathbf{x}^{(k)})\coloneqq D_{\textrm{KL}}(\hat{\mathbf{x}}||\mathbf{x}^{(k+1)})-D_{\textrm{KL}}(\hat{\mathbf{x}}||\mathbf{x}^{(k)}) 
\end{equation}
implies the ISS state. \textcolor{black}{The discrete-time derivative measures the change in relative entropy between an evolving state and an equilibrium (fixed) state. Notice,} 
\begin{subequations}
\label{eq:dklid}
\begin{align}
    \Delta D_{\textrm{KL}}(\hat{\mathbf{x}}||\mathbf{x}^{(k)})&=-\sum_{i=1}^2\hat{x}_i\ln{\left(x_i^{(k+1)}/x_i^{(k)}\right)}\\&=-\sum_i^2\hat{x}_i\ln{\left[1+(\delta_\varphi)_i^{(k)}\right]}\\&\geq
-\ln{\left[1+\sum_i^2\hat{x}_i(\delta_\varphi)_i^{(k)}\right]}.
\end{align} 
\end{subequations}
The last inequality is obtained using Jensen's inequality~\cite{Jensen1906,Cover2005} for convex functions. Now, for decreasing KL-divergence, we should have $\Delta D_{\textrm{KL}}(\hat{\mathbf{x}}||\mathbf{x}^{(k)})< 0$, or, $\sum_i\hat{x}_i(\delta_\varphi)^{(k)}_i > 0$. This implies,  $\sum_i\varphi_i(\mathbf{x}^{(k)})-\sum_i \hat{x}_i \left(\varphi_i(\mathbf{x}^{(k)})/x_i^{(k)}\right)< 0$  for all $\mathbf{x}^{(k)}$ in the deleted neighbourhood of $\hat{\mathbf{x}}$, which is the ISS condition. Note that if one could interpret the KL-divergence as the discrete-time Lyapunov function~\cite{Nicoletta_2018,wiggins_2003} for this dynamic, this result shows that the stable fixed point implies ISS for two-strategy games, which is in line with the result known using linear stability analysis. \textcolor{black}{Furthermore, an extension to higher $(n>2)$ strategy games is immediate: the superscripts in the summations can be replaced by $n$ in Eq.~\eqref{eq:dklid}.}

\textcolor{black}{For the results on information-theoretic interpretations of stability conditions in evolutionary dynamics in the subsequent parts of the paper, we will follow a line of reasoning similar to Eq.~\eqref{eq:dklid}, with the details of the mathematical steps relegated to the Appendices.} 
\subsection{Escort-Incentive Dynamics}
There is yet another way the replicator dynamic can be generalized. Specifically, consider the discrete $q$-deformed replicator dynamic~\cite{harper_escort} given by
 \begin{equation}
      \Delta x_i^{(k)}=(x_i^{(k)})^q(f_i(\mathbf{x}^{(k)})-\langle f(\mathbf{x}^{(k)})\rangle_q )\label{eq:qre}
\end{equation}
where $q$ is a positive real number and $\langle f(\mathbf{x}^{(k)})\rangle_q:= \sum_{j}(x_j^{(k)})^qf_j(\mathbf{x}^{(k)})/\sum_{j}(x_j^{(k)})^q$ is the $q$-generalized mean. Clearly, setting $q\rightarrow1$ reduces this dynamic to the standard replicator dynamic. Eq.~(\ref{eq:qre}) is, in turn, a special case of a larger set of dynamics given by the following discrete-time escort-incentive dynamic~\cite{harper_timescales}:
\begin{equation}\label{escort-incentive_map}
    \Delta x_i^{(k)}=\varphi_i(\mathbf{x}^{(k)})-\tilde{ \sigma}_i(\mathbf{x}^{(k)})\sum_j\varphi_j(\mathbf{x}^{(k)}),
\end{equation}
where  the escort distribution vector, $\bm{\tilde{\sigma}}(\mathbf{x}^{(k)})=(\tilde{ \sigma}_1(\mathbf{x}^{(k)}),\tilde{ \sigma}_2(\mathbf{x}^{(k)}))$ is defined as $(\sum_{i=1}^2\sigma(x_i))^{-1}\left(\sigma(x_1),  \sigma(x_2)\right)$. \textcolor{black}{Nondecreasing and strictly positive on $(0,1)$, the escort function, $\sigma$ (appropriately normalized), maps a discrete probability distribution into itself. In our context, the escort distributions, which are often relevant in nonextensive statistical systems \cite{beck1995thermodynamics}, help further generalize the notion of mean---used, e.g., in defining mean fitness of a population}. Different choices for different escort and incentive functions correspond to different dynamics, such as replicator, $q$-deformed replicator, projection, and exponential escort (see Table~\ref{tab:escort}). It is interesting to note that this generalization of the replicator dynamic is actually motivated through the information-theoretic framework~\cite{harper_timescales}. \textcolor{black}{In fact, Eq.~\eqref{incentive_map} is a special case of Eq.~\eqref{escort-incentive_map} (when $\sigma_i(\mathbf{x})=x_i$).} 

\begin{table}
\caption{\label{tab:escort}Escort and incentive function combinations for various evolutionary dynamics.}
\begin{ruledtabular}
\begin{tabular}{ccc}
Evolutionary dynamic&Incentive ($\varphi_i(\mathbf{x})$) & Escort ($\sigma_i(\mathbf{x})$)\\
\hline
Replicator& $x_if_i(\mathbf{x})$& $x_i$\\
Replicator with selection &$\beta x_if_i(\mathbf{x})$\footnote{Here, $\beta$ is the intensity of selection in the dynamic.}& $\beta x_i$\\
$q$-deformed replicator&$x_i^qf_i(\mathbf{x})$ & $x_i^q$\footnote{Here, $q=1$ will give the replicator dynamic.}\\
Exponential escort &$e^{x_i}f_i(\mathbf{x})$& $e^x$
  \\
\end{tabular}
\end{ruledtabular}
\end{table}

It is natural that the concepts of ESS and ISS can be further extended for the escort-incentive dynamics: A state of the population is an escort incentive stable state (EISS)~\cite{harper_timescales}, $\hat{\mathbf{x}}$, if for any $\mathbf{x}$ in the local deleted neighbourhood of $\hat{\mathbf{x}}$, we have 
\begin{equation}
      \sum_i \hat{x}_i \frac{\varphi_i(\mathbf{x})}{\sigma_i(\mathbf{x})}>\sum_ix_i\frac{\varphi_i(\mathbf{x})}{\sigma_i(\mathbf{x})}.\label{eq:EISS}
\end{equation}
Again, for the two-strategy escort-incentive map, using linear stability analysis, it is an easy exercise to show---as a generalization of the corresponding result of replicator map~\cite{pandit}---that a locally asymptotically stable fixed point of the two-strategy escort-incentive map is an EISS.   

It stands proven in the literature~\cite{harper_escort} that the escort divergence is the Lyapunov function for the continuous escort-replicator equation (which is an alternate escort distribution motivated evolutionary dynamic). We recall that the escort divergence, $D_{\sigma}(\mathbf{x}||\mathbf{x}')$---a generalization of the KL-divergence---is given by $D_{\sigma}(\mathbf{x}||\mathbf{x}')\coloneqq\sum_{i}\int_{x'_i}^{x_i} \left(\log_\sigma (v)- \log_\sigma (x'_i)\right) \textrm{d}v$, where $\log_\sigma (x)\coloneqq\int_1^x{\sigma(u)}^{-1}\textrm{d}u$ is called the escort logarithm. Motivated by this result, it is natural to check what negative discrete-time derivative of the escort divergence, defined as $\Delta D_{\sigma}(\hat{\mathbf{x}}||\mathbf{x}^{(k)})\coloneqq D_{\sigma}(\hat{\mathbf{x}}||\mathbf{x}^{(k+1)})-D_{\sigma}(\hat{\mathbf{x}}||\mathbf{x}^{(k)})$, implies and how it is connected with a fixed point, $\hat{\mathbf{x}}$, that is an EISS. \textcolor{black}{Here, demanding $\Delta D_{\sigma}(\hat{\mathbf{x}}||\mathbf{x}^{(k)})< 0$ implies
\begin{equation}\label{escort-incentive-fixed point}
   \sum_i \hat{x}_i \frac{\varphi_i(\mathbf{x})}{\sigma_i(\mathbf{x})}>\sum_ix_i\frac{\varphi_i(\mathbf{x})}{\sigma_i(\mathbf{x})} -C_\sigma(\hat{\textbf{x}},k) 
\end{equation}
where $C_{\sigma}(\mathbf{\hat{x}}, k)$ is a non-negative correction term (specific to two types) that arises because of the choice of bounds of the integrals. The details of this computation and the explicit form of $C_{\sigma}(\mathbf{\hat{x}}, k)$ are provided in Appendix~\ref{escort-incentive_fixedpoint_appendix}. Since the correction term is non-negative, the obtained condition is automatically satisfied by the EISS condition in Eq.~(\ref{eq:EISS}).}

In summary, we have understood how the decrease in an information-theoretic divergence, namely, the escort divergence (of which KL-divergence is a special case) corresponds to the fixed points that are ESS, ISS, and EISS of a wide class of discrete-time evolutionary dynamics. 
\section{Interpreting Periodic orbits} \label{sec:periodic_orbits}
Non-convergent outcomes, like periodic orbits, are quite common in evolutionary dynamics. In the discrete-time dynamics discussed in the preceding section, such outcomes~\cite{vilone, pandit,umezuki_logitperiodic, archan_periodic} are possible even with only two-strategy games, thus rendering the investigations about them analytically tractable. If we represent a finite sequence of $m$ states representing iterates of a map by $\{\mathbf{x}^{(k)}\}_{k=1}^{k=m}$ (where $k$ denotes the time step), then a periodic orbit of orbit $m$ (i.e., an $m$-period orbit) can be denoted by an
infinite sequence $\{\mathbf{x}^{(k)}\}_{k\geq 1}$  such that
$\mathbf{x}^{(k)}=\mathbf{x}^{(k+m)}$, for any $k\geq 1$, \textcolor{black}{where $m$ is the least (also known as prime) period of the orbit.} Since a periodic
orbit only has $m$ \textcolor{black}{distinct} states, it may be compactly denoted by a finite
sequence given by $\{\mathbf{x}^{(k)}\}_{k=1}^{k=m}$---a time-ordered collection of $m$-period points. Now we intend to show that the information-theoretic concept of KL-divergence (or more generally the escort divergence) presents a unique viewpoint that helps to construct a game-theoretic interpretation for the non-convergent dynamical equilibrium, specifically periodic orbit. 

It is imperative to bring to the readers' attention that an exercise of imparting game-theoretic meaning to the periodic orbits was performed in Ref.~\cite{archan_periodic}. We first critically revisit that work and scrutinize the concepts related to the periodic orbits within the game-theoretic framework. We confine ourselves only to the standard replicator map in this discussion.

\subsection{Scrutiny of Heterogeneity Stable Orbit}\label{scrutiny_hso}
Two definitions were introduced~\cite{archan_periodic} \textcolor{black}{as extensions of Nash equilibrium and evolutionarily stable strategy, namely, heterogeneity orbit (HO) and heterogeneity stable orbit (HSO). We formally define them below.} 

\textcolor{black}{\textit{Definition 1a (HO for replicator map)}.  A sequence of $m$ distinct states $\{\hat{\bf x}^{(k)}\}_{k=1}^{k=m}$ is an HO of order $m$ if for all states $\hat{\mathbf{x}}^{(j)}$ $(1\leq j\leq m)$ of this sequence,} 
 \begin{equation}
\sum_{k=1}^{m} H_{{\hat{\bf x}}^{(k)}} \Big[\hat{\mathbf{x}}^{(j)}\cdot f(\mathbf{\hat{x}}^{(k)})\Big] = \sum_{k=1}^{m} H_{\hat{\bf x}^{(k)}} \Big[\mathbf{x}\cdot f(\mathbf{\hat{x}}^{(k)})\Big].
\label{eq: HE v1} 
\end{equation}
Here, heterogeneity factor, $H_{\mathbf{x}}\equiv2x(1-x)$,  is the probability that two arbitrarily chosen members of the population belong to two different types. The condition of HO is equivalent to the condition of periodicity given by 
\begin{equation}\label{periodicity_condition}
    \sum_{j=k}^{k+m-1}\mathbf{\hat{x}}^{(j)}(1-\mathbf{\hat{x}}^{(j)})\left[f_1(\mathbf{\hat{x}}^{(j)})-f_2(\mathbf{\hat{x}}^{(j)})\right]=0,
\end{equation}
where $\mathbf{\hat{x}}^{(k)}$ ($1\leq k\leq m$), is any one of the $m$ states of an $m$-periodic orbit. For $m=1$, it reduces to the condition of the Nash equilibrium---$\mathbf{\hat{x}}\cdot f(\mathbf{\hat{x}})=\mathbf{x}\cdot f(\mathbf{\hat{x}})$ for any $\mathbf{x}$ in the interior of the simplex. 

\textcolor{black}{\textit{Definition 1b (HSO for replicator map)}. An HSO of order $m$ is a sequence of $m$ distinct states, $\{\hat{\bf x}^{(k)}\}_{k=1}^{k=m}$ such that}
\begin{equation}
\sum_{k=1}^{m} H_{{\bf{x}}^{(k)}} \big[\hat{\mathbf{x}}^{(1)}\cdot f(\mathbf{{x}}^{(k)})\Big] > \sum_{k=1}^{m} H_{{\bf{x}}^{(k)}} \big[{\mathbf{x}}^{(1)}\cdot f(\mathbf{{x}}^{(k)})\Big] ,
\label{eq:hso}
\end{equation}
\textcolor{black}{for any sequence of states $\{{\bf x}^{(k)}\}_{k=1}^{k=m}$ of the map starting in some infinitesimal deleted neighbourhood of $\hat{{\bf x}}^{(1)}$.}  The HSO reduces to ESS for $m=1$ and extends the concept of ESS to periodic orbits. 

Using linear stability, one can show that all locally asymptotically stable periodic orbits obey the HSO criterion, although the converse is not necessarily true~\cite{archan_periodic}. Moreover, HSO implies HO, just as ESS implies Nash equilibrium. To truly interpret the condition game-theoretically, however, a connection to  underlying strategy space is necessary; and in this context, it can be shown that just as the ESS condition is associated with the idea of strong stability~\cite{Hofbauer1998}, HSO may also be associated with the extension of strong stability for periodic orbits~\cite{archan_periodic}.  

Two criticisms of the definition of the HSO are in order. First, the introduction of the heterogeneity factor is somewhat ad hoc. Second, a close inspection reveals that the HSO condition is a bit odd in the following sense: The total payoff on the right hand side of Eq.~(\ref{eq:hso}) is not the sum of the evolving mutant state playing with itself, but instead is the sum of payoffs of the mutant playing against a temporally fixed mutant state chosen [note the fixed $(1)$ superscript on the right hand side of Eq.~(\ref{eq:hso})]. In this paper, we circumvent the aforementioned two drawbacks by introducing a new generalization of the ESS.    

\subsection{Information Stable Orbit}\label{subsec:iso}
Let us generalize the concept of ESS to define information stable orbit (ISO) as follows:

\textcolor{black}{\textit{Definition 2a (ISO for replicator map)}. A sequence of $m$ distinct states, $\{\hat{\bf x}^{(k)}\}_{k=1}^{k=m}$, is ISO of order $m$ of the replicator map if}
\begin{equation}
\label{ISO_replicator}
\sum_{k=1}^{m}\hat{\mathbf{x}}^{(1)}\cdot f(\mathbf{x}^{(k)})> \sum_{k=1}^{m}\mathbf{x}^{(k)}\cdot f(\mathbf{x}^{(k)}),
\end{equation}
\textcolor{black}{for any sequence of $m$ states, $\{{\bf x}^{(k)}\}_{k=1}^{k=m}$, of the map starting in some infinitesimal deleted neighbourhood of $\hat{{\bf x}}^{(1)}$.}

The definition of the ISO of order $m$ qualifies any given finite sequence of $m$ states of a map, $\{\mathbf{\hat{x}}^{(k)}\}_{k=1}^{k=m}$, by comparing it---using Eq.~(\ref{ISO_replicator})---with a finite sequence of $m$ mutant states, $\{\mathbf{x}^{(k)}\}_{k=1}^{k=m}$, also governed by the map. In doing so, only the first element of the given sequence, $\mathbf{\hat{x}}^{(1)}$, is explicitly used in the inequality, with the implicit knowledge that the remaining elements of this sequence are specified by the map exactly. The sequence for mutant states starts from the deleted neighbourhood of the first element of the ISO, and the remaining elements of this sequence are also determined exactly by the map. This formulation of a stability criterion for periodic orbits is inspired \textcolor{black}{by} the previously defined HSO~\cite{archan_periodic}. 

\textcolor{black}{Note that explicitly, Eq.~\eqref{ISO_replicator} is blind to the notion of a periodic orbit. Thus, in order to qualify a periodic orbit by ISO, we must ensure that each of the $m$ sequences of length $m$ starting from each of the $m$ distinct states of an $m$-periodic orbit should obey the ISO condition.} Here, all the $m$ sequences $\{\hat{\mathbf{x}}^{(j)}\}_{j=k}^{j=k+m-1}$ equivalently denote the same $m$-periodic orbit and corresponding nearby sequences $\{{\mathbf{x}}^{(j)}\}_{j=k}^{j=k+m-1}$ start in the infinitesimal deleted neighbourbood of $\hat{\mathbf{x}}^{(k)}$. {Hence, we can recast the ISO condition (Eq.~(\ref{ISO_replicator})) for periodic orbits as}
\begin{equation}\label{iso_for_periodicorbits}
     \sum_{j=k}^{k+m-1}\hat{\mathbf{x}}^{(k)}\cdot f(\mathbf{x}^{(j)})> \sum_{j=k}^{k+m-1}\mathbf{x}^{(j)}\cdot f(\mathbf{x}^{(j)}),
\end{equation}
for all $m$ states $\hat{\mathbf{x}}^{(k)}$ of the sequence $\{\hat{\mathbf{x}}^{(k)}\}_{k=1}^{k=m}$. Note that the ISO reduces to the ESS for $m=1$, which is consistent with the fact that a fixed point can be understood as a 1-period orbit.

Since any state of an $m$-period orbit of a nonlinear map is a fixed point of the $m$-th iterate of the map, it is very natural to ask if the condition of decrease in the KL-divergence leads to an extension of ESS, either HSO or ISO; after all, we have seen earlier that $\Delta D_{\textrm{KL}}(\hat{\mathbf{x}}||\mathbf{x}^{(k)})< 0$ leads to ESS [see Eqs.~(\ref{eq:dklid})a--c]. To this end, we introduce the notation,
\begin{equation}
\Delta_{m} D_{\textrm{KL}}(\hat{\mathbf{x}}^{(k)}||\mathbf{x}^{(k)})\equiv D_{\textrm{KL}}(\hat{\mathbf{x}}^{(k)}||\mathbf{x}^{(k+m)})-D_{\textrm{KL}}(\hat{\mathbf{x}}^{(k)}||\mathbf{x}^{(k)})
\end{equation}
which measures the change in the KL-divergence over $m$ time steps. \textcolor{black}{Thus, it is a measure of the total change in the relative entropy over a time period of a periodic orbit of the replicator map. 
Using Jensen's inequalities as used in the fixed-point computation (see Eq.~\eqref{eq:dklid}), we find that demanding $\Delta_{m} D_{\textrm{KL}}(\hat{\mathbf{x}}^{(k)}||\mathbf{x}^{(k)})< 0$ implies the ISO condition as given in Eq.~\eqref{iso_for_periodicorbits} and not the condition of HSO. The details of this proof are provided in Appendix~\ref{ISOproof_replicator}.} 

Thus, this information-theoretic fact justifies the aptness of the name ISO. One should observe the unidirectionality in the arguments \textcolor{black}{presented in Appendix~\ref{ISOproof_replicator}.} In particular, just as there exist some unstable fixed points that are ESS in the discrete replicator map, we may also find some unstable periodic orbits that obey the ISO condition. This unidirectionality is also true for the HSO condition and appears to be an artefact of the temporal discretization. 

As can be guessed, the generalization of the ISO for the incentive dynamics is straightforward:

\textcolor{black}{\textit{Definition 2b (ISO for incentive map)}. A sequence of $m$ distinct states, $\{\hat{\bf x}^{(k)}\}_{k=1}^{k=m}$, is ISO of order $m$ of the incentive map if}
\begin{eqnarray}
\label{eq:IISO}
\sum_{k=1}^{m}\sum_i\hat{x}_i^{(1)}\frac{\varphi_i(\mathbf{x}^{(k)})}{x_i^{(k)}}> \sum_{k=1}^{m}\sum_i\varphi_i(\mathbf{x}^{(k)})
\end{eqnarray}
\textcolor{black}{for any sequence of states, $\{{\bf x}^{(k)}\}_{k=1}^{k=m}$, of the map starting in some infinitesimal deleted neighbourhood of $\hat{{\bf x}}^{(1)}$.}

We observe that {this condition reduces} to ISO on setting $\varphi(\mathbf{x})=xf(\mathbf{x})$ as expected for the replicator map. Also, $m=1$ reproduces the ISS condition (Eq.~(\ref{eq:iss})) as required. \textcolor{black}{We remark that for the condition in Eq.~\eqref{eq:IISO} to hold for periodic orbits, we can recast it into a more general form similar to Eq.~\eqref{iso_for_periodicorbits}.} Then, just like for the ISO of the replicator map,  here one can show that the decreasing KL-divergence constructed using a periodic orbit and its neighbouring orbit implies the {ISO criterion for incentive map}---to prove so, all one has to do is to replace $\delta_{xf}$ by more general $\delta_{\varphi}$ in the \textcolor{black}{proof in Appendix~\ref{ISOproof_replicator}.}  

As in the discussion for fixed points, we now generalize ISO for the escort-incentive dynamic. 

\textcolor{black}{\emph{Definition 2c (ISO for escort-incentive map)}. A sequence of $m$ distinct states, $\{\hat{\bf x}^{(k)}\}_{k=1}^{k=m}$, is ISO of order $m$ of the escort-incentive map if}
\begin{eqnarray}
\label{eq:IEISO}     \sum_{k=1}^{m}\sum_i\hat{x}_i^{(1)}\frac{\varphi_i(\mathbf{x}^{(k)})}{\sigma_i(\mathbf{x}^{(k)})}> \sum_{k=1}^{m}\sum_ix_i^{(k)}\frac{\varphi_i(\mathbf{x}^{(k)})}{\sigma_i(\mathbf{x}^{(k)})}
\end{eqnarray}
\textcolor{black}{for any sequence of states, $\{{\bf x}^{(k)}\}_{k=1}^{k=m}$, of the map starting in some infinitesimal deleted neighbourhood of $\hat{{\bf x}}^{(1)}$.}

\textcolor{black}{The condition in Eq.~\eqref{eq:IEISO} can also be recast to a form like in Eq.~\eqref{iso_for_periodicorbits} for periodic orbits.} Here, in the context of the information-theoretic interpretation of {this condition}  that is an $m$-periodic orbit, the appropriate information-theoretic concept is the escort divergence. Hence, we introduce the notation, $\Delta_{m} D_{\sigma}(\hat{\mathbf{x}}^{(k)}||\mathbf{x}^{(k)})\coloneqq D_{\sigma}(\hat{\mathbf{x}}^{(k)}||\mathbf{x}^{(k+m)})-D_{\sigma}(\hat{\mathbf{x}}^{(k)}||\mathbf{x}^{(k)})$ which measures the change in the escort divergence over $m$ time steps. \textcolor{black}{We find that demanding $\Delta_{m}D_{\sigma}(\hat{\mathbf{x}}^{(k)}||\mathbf{x}^{(k)})<0$, implies
\begin{eqnarray}
	\label{eqn:ISO_general}
	\sum_{j=k}^{k+m-1}\sum_i\hat{x}_i^{(k)}\frac{\varphi_i(\mathbf{x}^{(j)})}{\sigma_i(\mathbf{x}^{(j)})}> \sum_{j=k}^{k+m-1}\sum_ix_i^{(j)}\frac{\varphi_i(\mathbf{x}^{(j)}}{\sigma_i(\mathbf{x}^{(j)})}\nonumber\\
	-\sum_{j=k}^{k+m-1}C_\sigma(\hat{\textbf{x}}^{(k)},j).~~
\end{eqnarray}
where $C_\sigma(\hat{\textbf{x}}^{(k)},j)$ is the non-negative correction term introduced in Eq.~\eqref{escort-incentive-fixed point}, which arose due to the choice of bounds in the computation. The derivation of this result is given in Appendix~\ref{appendix:escort_periodic}.} We observe that if a periodic orbit satisfies the definition of ISO for escort-incentive map (Eq.~(\ref{eq:IEISO})), then it automatically satisfies the immediately preceding inequality~(\ref{eqn:ISO_general}). Hence, the relation of the ISO for escort-incentive map, that is a periodic orbit, with decreasing escort divergence is transparent.

\subsection{\textcolor{black}{Dynamical Stability and Game-Theoretic Interpretation}}\label{sec:dyn_stab}

\textcolor{black}{In subsection~\ref{subsec:iso}, we have motivated the ISO condition from information-theoretic grounds. We now illustrate the central idea of the paper through a proposition that brings together the three perspectives---information-theoretic, game-theoretic, and dynamic---whose interconnection is what this paper is all about.} 

\textcolor{black}{For simplicity, we consider the replicator map.} As described in subsection~\ref{scrutiny_hso}, we note that a locally asymptotically stable periodic orbit of (two-player two-strategy) replicator map is an HSO \cite{archan_periodic}. Does such a fact exist between periodic orbit and ISO as well? We find that the answer is affirmative (although, the mathematical proof is elusive for periodic orbit of general period). \textcolor{black}{We provide the following result for a two-period orbit:} 
 
\textit{Proposition 1}. A locally asymptotically stable 2-period orbit of a replicator map for two-player two-strategy game is an ISO.

\textit{Proof}. Since we know that a locally asymptotically stable 2-periodic orbit must be HSO~\cite{archan_periodic}, \textcolor{black}{we rewrite the HSO condition (Eq.~(\ref{eq:hso})) as follows}
\begin{align}
    \nonumber &\sum_{j=1}^2\left[\hat{x}^{(1)}\cdot f(\mathbf{x}^{(j)})-x^{(j)}\cdot f(\mathbf{x}^{(j)})\right]>\\
    &\left[\frac{H_{\mathbf{x}^{(2)}}}{H_{\mathbf{x}^{(1)}}}\left((\mathbf{x}^{(1)}-\hat{\mathbf{x}}^{(1)})\cdot f(\mathbf{x}^{(2)})\right)-(\mathbf{x}^{(2)}-\hat{\mathbf{x}}^{(1)})\cdot f(\mathbf{x}^{(2)})\right].\label{eq:maan}
\end{align}
This way of writing makes the terms, required to define the ISO condition, appear conveniently together on the left hand side; hence, to show that the ISO condition holds, one has to simply show that the right hand side is nonnegative. Now, the right hand side may be recast as
\begin{align}
    \nonumber &\left[f_1(\mathbf{x}^{(2)})\!-\!f_2(\mathbf{x}^{(2)})\right]\!\!\left[\left(\hat{x}^{(1)}-x^{(2)}\right)\!-\!\frac{H_{\mathbf{x}^{(2)}}}{H_{\mathbf{x}^{(1)}}}\left(\hat{x}^{(1)}-x^{(1)}\right)\right]\\   &=\!2(H_{\mathbf{x}^{(2)}})^{-1}\Delta{x}^{(2)}\left[\left(\hat{x}^{(1)}-x^{(2)}\right)\!-\!\frac{H_{\mathbf{x}^{(2)}}}{H_{\mathbf{x}^{(1)}}}\left(\hat{x}^{(1)}-x^{(1)}\right)\right],\label{eq:naam}
\end{align}
where \textcolor{black}{$\Delta x^{(2)}\coloneqq x^{(3)}-x^{(2)}$ (as in the notation introduced in Eq.~(\ref{eq:ret1}))}. We are free to choose small enough neighbourhood of $\hat{\bf x}^{(1)}$ so that ${\bf x}^{(1)}$ is close enough such that the term containing $\hat{x}^{(1)}-x^{(1)}$ on the right-hand side of Eq.~(\ref{eq:naam}) can be made arbitrarily small (note that its prefactor composed of heterogeneity factors is finite). Furthermore, for stable periodic orbits and the aforementioned small neighbourhood, $\Delta{x}^{(2)}$ and $\hat{x}^{(1)}-x^{(2)}$ are of same sign. In conclusion, the right-hand side of  Eq.~(\ref{eq:maan}) is a positive quantity for some small neighbourhood of  $\hat{\bf x}^{(1)}$. Therefore, the right-hand side of  Eq.~(\ref{eq:maan}) is greater than zero, and the aforementioned proposition stands proven.

\textcolor{black}{Thus, Proposition 1 illustrates a connection between the dynamical stability of periodic outcomes and the ISO equilibrium. We conjecture that such results may also be derived for higher period orbits although it appears to be mathematically more involved and require the specific properties of the dynamic in question.}

\textcolor{black}{To appreciate the game-theoretic point-of-view, we observe how the game-theoretic idea of stability in terms of comparison between payoffs seamlessly translates to the stability criterion for periodic orbits as well. There is no explicit need for a heterogeneity-weighted payoff as considered in a prior work~\cite{archan_periodic}. The periodicity condition in Eq.~\eqref{periodicity_condition} can be treated as the extension of Nash equilibrium for periodic orbits. ISO serves as the extension of ESS. An ESS state is a state such that its payoff, when played against any other state in its neighbourhood, is greater than the payoff gained when the other state plays with itself. Analogously, for the ISO, we see that the relevant quantity for comparison is the {sum} of payoffs for a particular state $\hat{\mathbf{x}}$ of the $m$-period orbit. Here we must look at a sequence of $m$ states starting from the neighbourhood of the state $\hat{\mathbf{x}}$ and compare the sum of the payoffs---as gained by the state $\hat{\mathbf{x}}$ while in play with each of the states in the sequence---with the sum of the payoff that the states in the sequence gain while playing against themselves. In other words, we may say that the $m$-period orbit is evolutionarily stable when a state in the orbit outperforms an evolving mutant state as compared to when the mutant state keeps playing with itself.}

\textcolor{black}{Finally, we remark that the definitions of ISO (Definition 2a-c) are constructed from a game-theoretic perspective, without invoking any notion of (periodic) dynamics. Thus, although our entire analysis pertains to two-player two-strategy games, these definitions also hold for two-player $n$-strategy games (with $n>2$).} 

\subsection{Numerical Verification} \label{sec:numerical_results}
\begin{figure}[!t]
    \centering
    \includegraphics[width=\columnwidth]{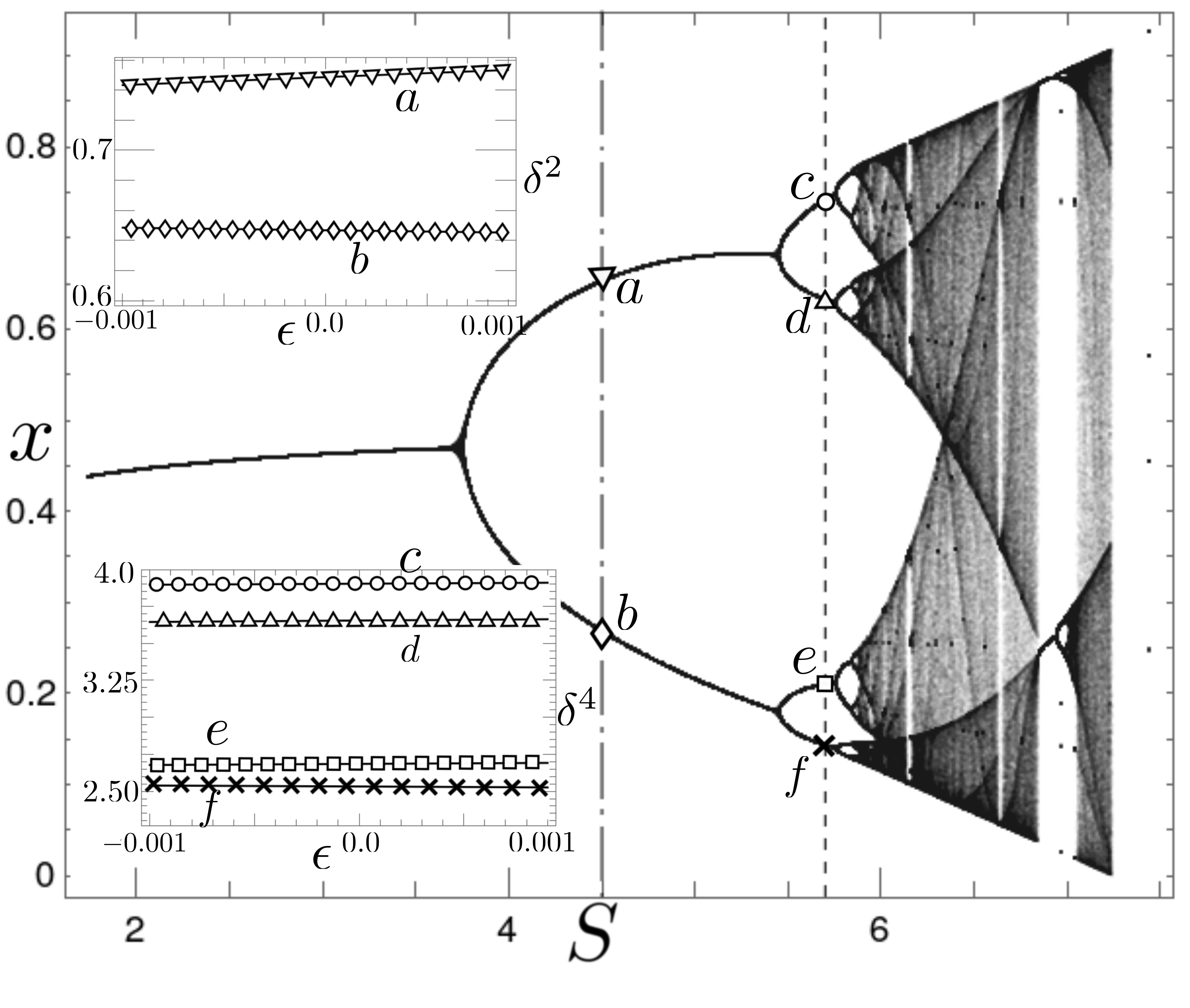}
    \caption{\textcolor{black}{\emph{Stable periodic orbits are ISO for replicator map:} The figure presents the bifurcation diagram for the replicator map for a two-player two-strategy game. The payoff matrix, ${\sf \Pi}$, is chosen such that $T=1.5+S$, $R=1$ and $P=0$. The dash-dotted line, $S=4.5$, intersects the bifurcation diagram at 2-period points---$a\approx0.73$ (inverted triangle) and $b\approx0.26$ (diamond)---of the 2-period orbit. The dashed line, $S=5.7$, cuts the bifurcation diagram at the 4-period points of the 4-period orbit---$\{c,d,e,f\}\approx\{0.74, 0.63, 0.20, 0.14\}$; the points are respectively marked by circle, triangle, square and cross. Upper inset: It depicts positivity of $\delta^2\equiv\delta^2_{\varphi,\sigma}(\hat{\mathbf{x}}^{(k)},\epsilon)$ [calculated for the stable 2-period points $a$ and $b$; see Eq.~(\ref{eq:fig1})] for small values of $\epsilon$. The curves for $\delta^2_{\varphi,\sigma}((a,1-a),\epsilon)$ vs. $\epsilon$ and $\delta^2_{\varphi,\sigma}((b,1-b),\epsilon)$ vs. $\epsilon$ are marked respectively by inverted triangles and diamonds. Lower inset: Similar to what is done in the upper inset, it showcases positivity of  $\delta^4\equiv\delta^4_{\varphi,\sigma}(\hat{\mathbf{x}}^{(k)},\epsilon)$ corresponding to  the $4$-period points---$c,\,d,\,e,\,$~and~$f$---represented by the curves respectively  marked by circles, triangles, squares and crosses.}}
    \label{fig:replicator}
\end{figure}

\begin{figure}[!t]
    \centering
    \includegraphics[width=\columnwidth]{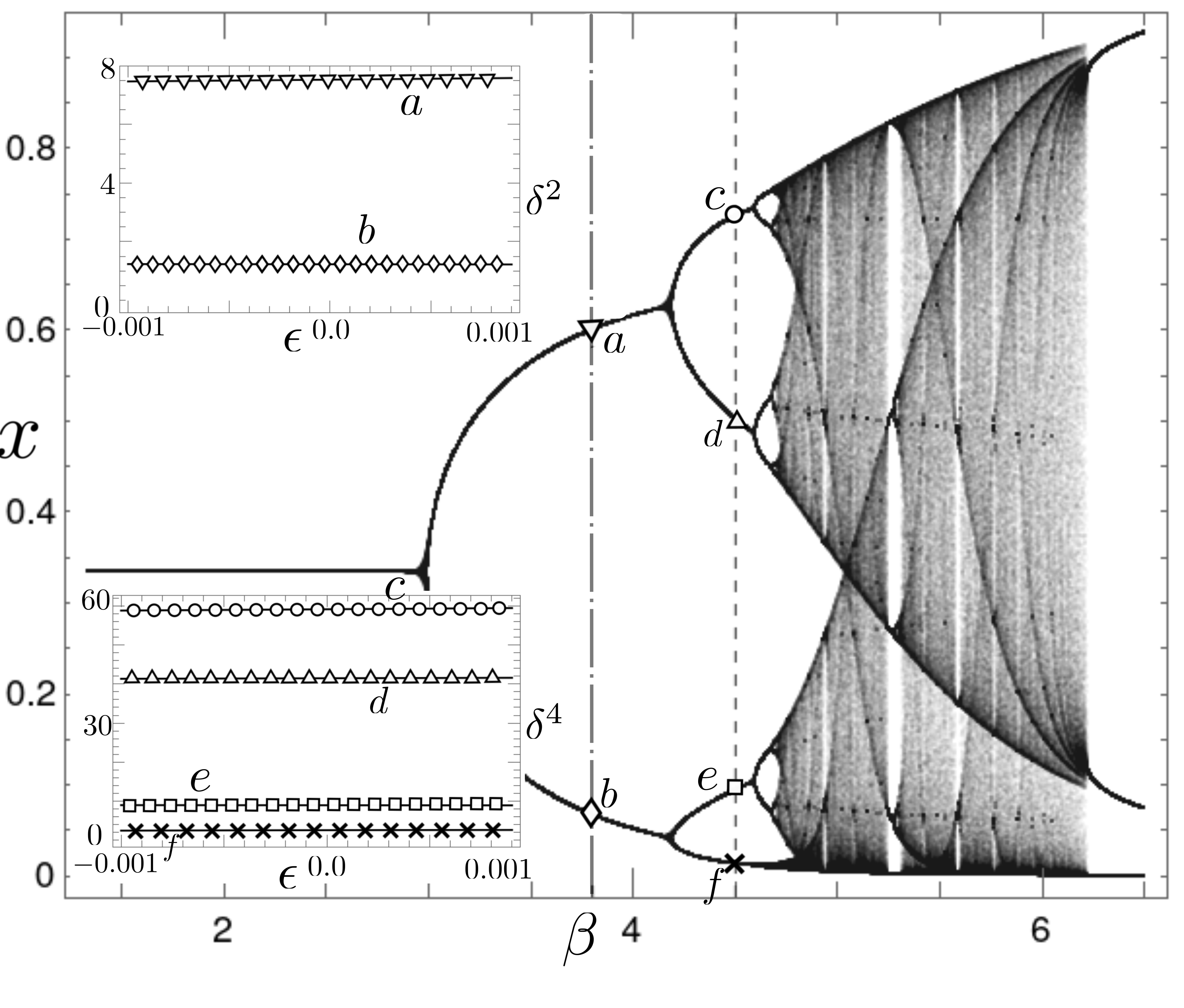}
      \caption{\textcolor{black}{\emph{Stable periodic orbits are ISO for incentive map:} The figure---analogous to Fig.~(\ref{fig:replicator})---presents the bifurcation diagram for the logit map for two-player two-strategy game. The payoff matrix, ${\sf \Pi}$, is chosen such that $R=1$, $S=2$, $T=3$, and $P=1$. Upper inset: {It depicts} positivity of $\delta^2\equiv\delta^2_{\varphi,\sigma}(\hat{\mathbf{x}}^{(k)}, \epsilon)$ for the two states of the stable 2-period orbit points---$a\approx0.59$ (inverted triangle) and $b\approx0.06$ (diamond)---corresponding to $\beta=3.8$  (see vertical dash-dotted line intersecting the bifurcation diagram). Lower inset: {It} shows positivity of  $\delta^4\equiv\delta^4_{\varphi,\sigma}(\hat{\mathbf{x}}^{(k)}, \epsilon)$ for the four states of the  $4$-period orbit, $\{c,d,e,f\}\approx\{0.72, 0.50, 0.10, 0.09\}$, {denoted by circle, triangle, square and cross respectively corresponding to $\beta=4.5$  (see vertical dashed line intersecting the bifurcation diagram).}}}
    \label{fig:logit}
\end{figure}
\begin{figure}[h]
    \centering
    \includegraphics[width=\columnwidth]{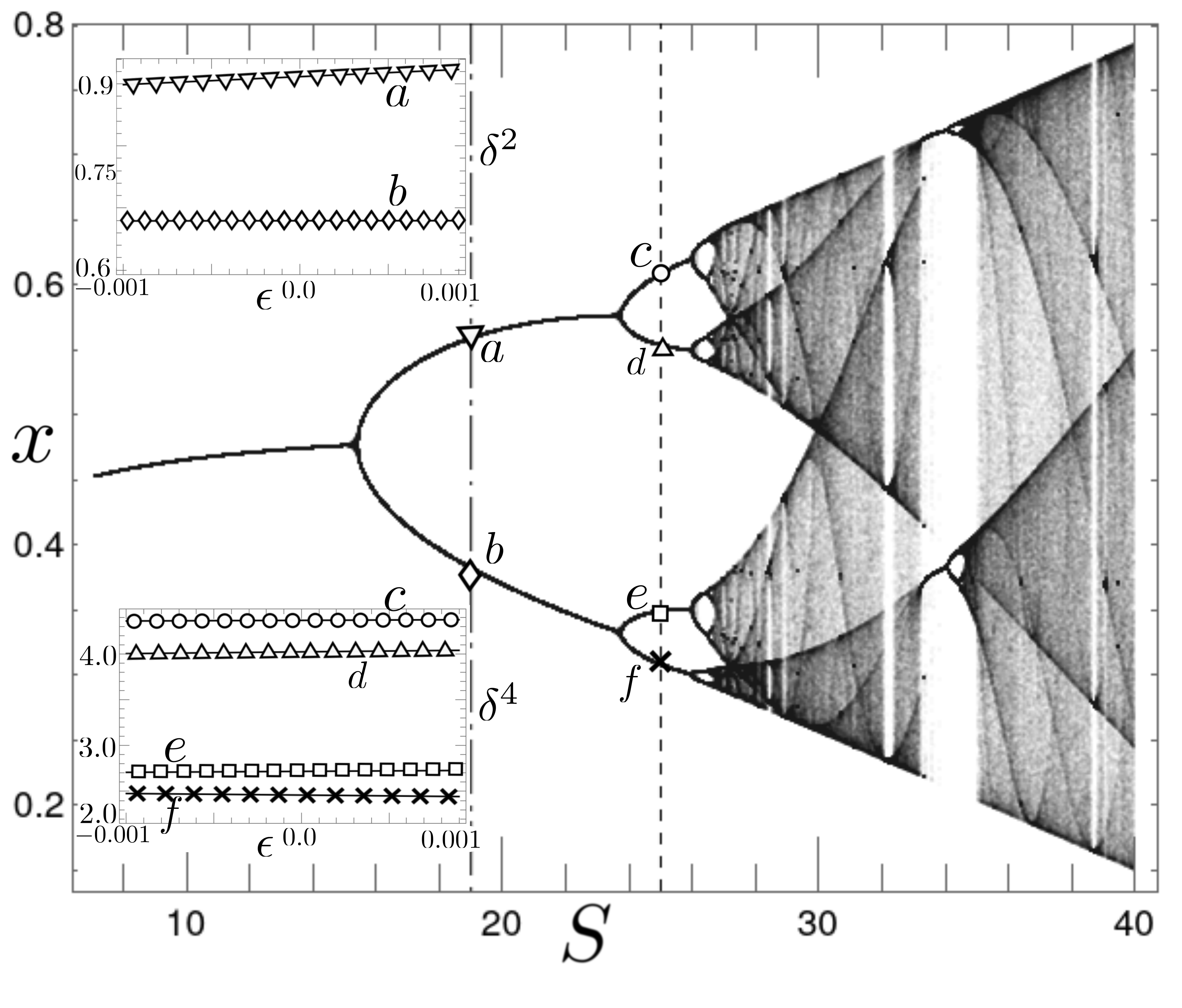}
          \caption{\textcolor{black}{\emph{Stable periodic orbits are ISO for escort-incentive map:} This figure---analogous to Fig.~(\ref{fig:replicator})---showcases the bifurcation diagram for the $q$-replicator map for two-player two-strategy games. The payoff matrix, ${\sf \Pi}$, is chosen such that $R=1$, $P=0$, and $T=2.5+S$; $q$ has been chosen to be $3$. Upper inset: It shows positivity of $\delta^2\equiv\delta^2_{\varphi,\sigma}(\hat{\mathbf{x}}^{(k)}, \epsilon)$ for the two states of the stable 2-period orbit points---$a\approx0.56$ (inverted triangle) and $b\approx0.36$ (diamond)---existing at $S=19.0$ (see vertical dash-dotted line intersecting the bifurcation diagram). Lower inset: It depicts positivity of $\delta^4\equiv\delta^4_{\varphi,\sigma}(\hat{\mathbf{x}}^{(k)}, \epsilon)$ for the four states of the $4$-period orbit, $\{c,d,e,f\}\approx\{0.60, 0.55, 0.34, 0.30\}$, denoted by circle, triangle, square and cross respectively corresponding to $S=25.0$ (see vertical dashed line intersecting the bifurcation diagram).}}
    \label{fig:qrep}
\end{figure}
We now undertake the exercise of numerically illustrating that the periodic orbits indeed are ISO or its generalizations {for the incentive and escort-incentive maps}.  For this purpose, it helps to denote an $m$-period orbit as $\{\hat{\bf x}^{(k)}\}_{k=1}^{k=m}$and so 
$\{{\bf x}^{(k)}\}_{k=1}^{k=m}$ is the neighbouring trajectory with $x^{(1)}=\hat{x}^{(1)}+\epsilon$. Furthermore, the {ISO condition for the escort-incentive map} ({it being the most general version}) may be compactly written as $ \delta^m_{\varphi,\sigma}(\hat{\mathbf{x}}^{(k)})>0$ where
\begin{equation}
    \delta^m_{\varphi,\sigma}(\hat{\mathbf{x}}^{(k)},\epsilon)\coloneqq\sum_{j=k}^{k+m-1}\sum_i\left[\hat{x}_i^{(k)}\frac{\varphi_i(\mathbf{x}^{(j)})}{\sigma_i(\mathbf{x}^{(j)})}- x_i^{(j)}\frac{\varphi_i(\mathbf{x}^{(j)})}{\sigma_i(\mathbf{x}^{(j)})}\right].\label{eq:fig1}
\end{equation}
We want to show that the condition is satisfied at least for some small values of $\epsilon$.

In addition to the replicator map, we take two more maps---logit map, and $q$-replicator map---to respectively represent the classes of incentive and escort-incentive maps. The fitness of $i$-th strategy (or type) $f_i(\mathbf{x})$, is calculated using a payoff matrix, 
\begin{equation}
\begin{tabular}{c}
 ${\sf{\Pi}}$ =
 $\begin{bmatrix}  
R & S \\ T & P
\end{bmatrix}$\, $(R,S,T,P\in\mathbb{R})$,
\end{tabular}\vspace{2mm}
\label{eqn:PayOff_A}
\end{equation}
so that $f_i(\mathbf{x})=({\sf{\Pi}}\mathbf{x})_i$. \textcolor{black}{It should be remarked that the elements of the payoff matrix must be chosen carefully so as to achieve forward-invariability of the corresponding maps, as noted in Sec.~\ref{fixed point} (see Ref.~\cite{pandit,MCC_JPC_21}).} 

For each of the dynamics, we first demonstrate the existence of periodic orbits by generating their bifurcation diagrams, and subsequently choose parameter values to obtain examples of 2-period and 4-period orbits. We then verify the positivity of $\delta^m_{\varphi,\sigma}(\hat{\mathbf{x}}^{(k)},\epsilon)$ for each of these examples using small values of $\epsilon$. The stability of the periodic orbit can be further verified using linear stability analysis and by observing the asymptotic behaviour of the corresponding time series. We systematically exhibit examples of stable $2$-period orbit ($\{a,b\}$) and $4$-period orbit ($\{c,d,e,f\}$) of replicator, logit and $q$-replicator dynamics  (see Figs.~\ref{fig:replicator},~\ref{fig:logit}~and~\ref{fig:qrep} respectively) to illustrate that they {follow the ISO criterion}.

\section{Discussion and Conclusions}
\label{sec:discussion}
Our results in this paper lie in the exciting overlapping area of evolutionary game theory, dynamical systems theory, and information theory. Specifically, we have shown that an extension of the idea of ESS is possible for periodic orbits in a large class of discrete-time evolutionary dynamics. We have termed the extended concept as ISO (and its generalizations) because it can be motivated through the information-theoretic idea of decreasing KL-divergence---a rather general principle that extends the idea of entropy maximization~\cite{Cover2005} in natural world. Thus, the concept of the ISO---in line with similar recent developments~\cite{baez_entropy,harper2010replicator,harper2009information}---highlights the tightly knit connections between the fields of evolutionary dynamics and information theory. These connections between evolutionary game dynamics and information theory is a promising avenue of research, and we suspect that a lot of information-theoretic concepts (such as R\'enyi entropy, Fisher information, and information geometry) may be transported to this context to improve our understanding of evolutionary systems. In particular, the possibility of an information-metric under which the discrete-time replicator map is a gradient-like system is worth pondering, and it could lead to tighter conditions of the stability for periodic orbits in game-theoretic terms.

From the game-theoretic view point, the ISO is a more satisfying extension of the ESS as compared to HSO~\cite{archan_periodic} because it does not require the use of the ad hoc heterogeneity factor in its definition. Furthermore, recall that ESS condition is tied to the concept of strong stability~\cite{Hofbauer1998} and it is not surprising that HSO is connected to the concept of the strongly stable strategy set (SSSS)~\cite{archan_periodic}---the extension of strong stability for periodic orbits. This is an important qualification, for the strong stability provides perhaps the `best validation for the concept of evolutionary stability'~\cite{Hofbauer1998}. Since locally asymptotically stable $2$-period orbit---being HSO---is ISO, when the orbit is SSSS, it is ISO as well. 

Before we end, we would like to delve into some related subtlety that deserves thorough future investigation. We wonder how the unstable periodic orbits connect with the ISO. We know that some unstable orbits are HSO and some are not~\cite{archan_periodic}. Although we are presently unable to provide any rigorous proof but we conjecture that a similar scenario is true for the case of ISO: A trivial example is that of the replicator map for two-player two-strategy stag-hunt game where there is an unstable interior fixed point ($1$-period orbit) that is not an ESS (or ISO of order 1)~\cite{pandit} (cf. Appendix~\ref{sec:A1}). 

\textcolor{black}{Since the replicator dynamic in a structured population remains same as the one for the unstructured population---only the payoff matrix gets modified~\cite{ohtsuki2006replicator,ohtsuki2008evolutionary} to include the effects of the structure---our conclusions should carry over to structured infinite populations.} To end with an optimistic note, we envisage useful extensions of the concepts developed in this paper for asymmetric games and extensive games, as well as for games in the finite populations. \textcolor{black}{However, for finite populations \cite{nowak2004emergence}, the notions of mixed ESS and existence of periodic orbits have to be defined before an information-theoretic perspective may be constructed.} Probably, the idea of ISO can be useful for the continuous-time evolutionary dynamics as well through the construction of Poincar\'e maps---an avenue worth pursuing in future.

\begin{acknowledgments}
\textcolor{black}{SB acknowledges the SURGE Programme at IITK (where part of the work was carried out) and support from the KVPY fellowship.} SC acknowledges the support from SERB (DST, govt. of India) through Project No. MTR/2021/000119.
\end{acknowledgments}
\appendix

\section{Escort-incentive stable state (EISS)}\label{escort-incentive_fixedpoint_appendix}
\textcolor{black}{In this appendix, we show how demanding decreasing relative entropy gives a condition consistent with the game-theoretic EISS condition for escort-incentive dynamics.    
For this computation, we need the following two inequalities obtained via usual bounds on Riemann integrals: 
\begin{subequations}
\begin{align}
\int_{a}^{b}\log_\sigma(v)\:\textrm{d}v&\geq (b-a)\log_\sigma(a),\label{eq:logs}\\
 \int_{a}^{b}\frac{1}{\sigma(u)}\textrm{d}u&\geq \frac{b-a}{\sigma(b)}.\label{eq:sig}
    \end{align}
    \end{subequations}
By making use of these inequalities, we compute the total change in relative entropy over a time period as follows. 
\begin{subequations}
\begin{eqnarray}
&&\Delta D_{\sigma}(\hat{\mathbf{x}}||\mathbf{x}^{(k)})=\sum_{i=1}^2\Big[-\left(\hat{x}_i-x_i^{(k+1)}\right)\log_\sigma(x_i^{(k+1)})\nonumber\\
&&+\left(\hat{x}_i-x_i^{(k)}\right)\log_\sigma (x_i^{(k)})+    \int_{x_i^{(k+1)}}^{x_i^{(k)}}\log_\sigma (v)\: \textrm{d}v\Big]\label{eq:9a}\\
    &&\geq \sum_{i=1}^2\left[\left(\hat{x}_i-x_i^{(k)}\right)\left( \log_\sigma (x_i^{(k)})- \log_\sigma (x_i^{(k+1)}) \right)\right]\label{eq:9b}\\ 
     &&= \sum_{i=1}^2\left[\left(\hat{x}_i-x_i^{(k)}\right) \int_{x_i^{(k+1)}}^{x_i^{(k)}}\frac{1}{\sigma(u)}\textrm{d}u\right] \label{int_sigma}\label{eq:csp}\\
     &&\nonumber \geq  \sum_{i}^2\left[\left(\hat{x}_i-x_i^{(k)}\right)\frac{\left(x_i^{(k)}-x_i^{(k+1)}\right)}{\sigma(x_i^{(k)})}\right]+\\
     &&(\hat{x}_2-x_2^{(k)})(x_2^{(k)}-x_2^{(k+1)})\!\!\left[\frac{1}{\sigma(x_2^{(k+1)})}-\frac{1}{\sigma(x_2^{(k)})}\right]\label{crucial step}\quad\\
    && =-\!\!\sum_{i}^2\hat{x}_i\frac{\varphi_i(\mathbf{x}^{(k)})}{\sigma(x_i^{(k)})}+\!\sum_{i}^2 x_i^{(k)}\frac{\varphi_i(\mathbf{x}^{(k)})}{\sigma(x_i^{(k)})}-\!C_\sigma(\hat{\textbf{x}},k),\label{eq:9e}
    \end{eqnarray}
    \end{subequations}
where, the correction term,  $C_\sigma(\hat{\textbf{x}},k)$, is given by 
\begin{align}\label{eq:correction_term}
\nonumber C_{\sigma}(\mathbf{\hat{x}}, k)\coloneqq &\left|\hat{x}_1-x_1^{(k)}\right|\left(x_2^{(k)}-x_2^{(k+1)}\right)\\
&\Big[{\sigma(x_2^{(k+1)})}^{-1}-{\sigma(x_2^{(k)})}^{-1}\Big]
\end{align}
The first inequality in Eq.~\eqref{eq:9b} is obtained by using the bound in Eq.~\eqref{eq:logs}, while the second inequality in Eq.~\eqref{crucial step} is obtained by using Eq.~\eqref{eq:sig}. However, to compare with the EISS expression, we notice that it is necessary to separate out a correction term $C_\sigma(\hat{\textbf{x}},k)$ given explicitly in Eq.~\eqref{eq:correction_term}. This expression is obtained from Eq.~\eqref{crucial step} by noting that $(\hat{x}_2-x_2^{(k)})=-(\hat{x}_1-x_1^{(k)})$ and that without any loss of generality, one can fix $\hat{x}_1-x_1^{(k)}>0$. Further, we note that since the escort function is nondecreasing, we have $(x_2^{(k)}-x_2^{(k+1)})\left[{\sigma(x_2^{(k+1)})}^{-1}-{\sigma(x_2^{(k)})}^{-1}\right]\ge0$. This implies that the $C_{\sigma}(\mathbf{\hat{x}}, k)$ is non-negative. Now, requiring the stability of fixed point, implies $\Delta D_{\sigma}(\hat{\mathbf{x}}||\mathbf{x}^{(k)})< 0$, which gives us Eq.~\eqref{escort-incentive-fixed point}, and that is consistently satisfied by the EISS condition (Eq.~(\ref{eq:EISS})).}

\section{Information stable orbit (ISO) for replicator map}\label{ISOproof_replicator}
\textcolor{black}{Here, we show how the requirement of decreasing total relative entropy leads to the ISO condition for the replicator map. We compute as follows.  
\begin{subequations}
\label{eq:dkliso}
\begin{align}
&\Delta_{m} D_{\textrm{KL}}(\hat{\mathbf{x}}^{(k)}||\mathbf{x}^{(k)})=-\sum_{i=1}^2\hat{x}_i^{(k)}\ln{\frac{x_i^{(k+m)}}{x_i^{(k)}}}\\
    &\qquad=-m\sum_i^2\hat{x}_i^{(k)}\sum_{j=k}^{k+m-1}\frac{1}{m}\ln{\left[1+(\delta_{xf})_i^{(j)}\right]}\\
    &\qquad\geq-m\sum_i^2\hat{x}_i^{(k)}\ln{\left[1+\sum_{j=k}^{k+m-1}\frac{1}{m}(\delta_{xf})_i^{(j)}\right]}\\
    &\qquad\geq -m\ln{\left[1+\sum_i^2\hat{x}_i^{(k)}\sum_{j=k}^{k+m-1}\frac{1}{m}(\delta_{xf})_i^{(j)}\right]},
\end{align}
\end{subequations}
where the last two inequalities are obtained using Jensen's inequality. One finds that demanding $\Delta_{m} D_{\textrm{KL}}(\hat{\mathbf{x}}^{(k)}||\mathbf{x}^{(k)})< 0$ implies the ISO condition as given in Eq.~(\ref{iso_for_periodicorbits}). Also note that an extension to higher strategy space may be possible here too (as in for the fixed point analysis in Eq.~\eqref{eq:dklid}), we simply replace the superscripts on the summations by $n$ ($n>2$).}

\section{Information stable orbit (ISO) for escort-incentive map}\label{appendix:escort_periodic}
\textcolor{black}{Here, we show how the requirement of decreasing total relative entropy leads to a condition consistent with the ISO condition for the escort-incentive map. We compute this by observing that the total relative entropy can be written as a telescoping sum of intermediate differences of relative entropies. Thus,  
\begin{subequations}
\begin{eqnarray}
\nonumber &&\Delta_{m}D_{\sigma}(\hat{\mathbf{x}}^{(k)}||\mathbf{x}^{(k)})\\
    &&=\sum_{j=k}^{k+m-1}\left[D_{\sigma}(\hat{\mathbf{x}}^{(k)}||\mathbf{x}^{(j+1)})-D_{\sigma}(\hat{\mathbf{x}}^{(k)}||\mathbf{x}^{(j)})\right]\\
    &&=\sum_{j=k}^{k+m-1}\Delta D_{\sigma}(\hat{\mathbf{x}}^{(k)}||\mathbf{x}^{(j)})\\
    &&\nonumber \geq -\sum_{j=k}^{k+m-1}\sum_{i=1}^2\hat{x}_i^{(k)}\frac{\varphi_i(\mathbf{x}^{(j)})}{\sigma(x_i^{(j)})}+\sum_{j=k}^{k+m-1}\sum_{i=1}^2x_i^{(j)}\frac{\varphi_i(\mathbf{x}^{(j)})}{\sigma(x_i^{(j)})}\\
    &&\quad-\sum_{j=k}^{k+m-1}C_\sigma(\hat{\textbf{x}}^{(k)},j).
\end{eqnarray}
\end{subequations}
The last inequality is obtained by using Eq.~(\ref{eq:9e}) and hence $\sum_{j=k}^{k+m-1}C_\sigma(\hat{\textbf{x}}^{(k)},j)$---a sum of nonnegative quantities---is nonnegative. This means that if the total escort divergence is decreasing, i.e., $\Delta_{m}D_{\sigma}(\hat{\mathbf{x}}^{(k)}||\mathbf{x}^{(k)})<0$, then following is implied:
\begin{eqnarray}
\label{ISO_general}
    \sum_{j=k}^{k+m-1}\sum_i\hat{x}_i^{(k)}\frac{\varphi_i(\mathbf{x}^{(j)})}{\sigma_i(\mathbf{x}^{(j)})}> \sum_{j=k}^{k+m-1}\sum_ix_i^{(j)}\frac{\varphi_i(\mathbf{x}^{(j)}}{\sigma_i(\mathbf{x}^{(j)})}\nonumber\\
    -\sum_{j=k}^{k+m-1}C_\sigma(\hat{\textbf{x}}^{(k)},j),\,\,
\end{eqnarray}
which is consistent with the ISO condition given in Eq.~\eqref{eq:IEISO}. 
}

\section{A note on 2-period orbit and ISO}
\label{sec:A1}
Here we demonstrate a curious result for the analytically tractable case of $2$-period orbits in the replicator map. In particular, we prove the following.

\textit{Proposition 2}. A two-period orbit of the replicator map for two-player two-strategy game is an ISO.

\textit{Proof}. Consider a periodic orbit $\{\hat{\textbf{x}}^{(1)}, \hat{\textbf{x}}^{(2)}\}$ and note that, by definition, $\Delta \hat{x}^{(1)}\Delta \hat{x}^{(2)}<0$. If we consider the two points of the periodic orbit to be well-separated and a state ${\textbf{x}}^{(1)}$ in the infinitesimal small neighbourhood of $\hat{\textbf{x}}^{(1)}$, then we expect $\Delta {x}^{(1)}\Delta {x}^{(2)}<0$---even if the periodic orbit is unstable, as long as ${\textbf{x}}^{(2)}$ (the state to which ${\textbf{x}}^{(1)}$ is mapped) is in small neighbourhood of $\hat{\textbf{x}}^{(2)}$, far away from ${\textbf{x}}^{(1)}$. We should always be able to write
\begin{eqnarray}
    \Delta x^{(1)}\Delta x^{(2)}+\sum_{j={1}}^2\left(\frac{H_{\mathbf{x}^{(2)}}}{H_{\mathbf{x}^{(j)}}}\right)\Delta x^{(j)}&\left(x^{(1)}-\hat{x}^{(1)}\right)<0\qquad
\end{eqnarray}
because the second term in the left hand side can be made infinitesimally small by taking $x^{(1)}\rightarrow \hat{x}^{(1)}$ as the heterogeneity factor is a positive quantity. Rewriting this equation, we arrive at 
\begin{equation}   \sum_{j=1}^2\left(H_{\mathbf{x}^{(j)}}\right)^{-1}\Delta x^{(j)}(x^{(j)}-\hat{x}^{(1)})<0,
\end{equation} 
which may be further rewritten using the replicator map to yield
\begin{equation}
    \sum_{j=1}^2(x^{(j)}-\hat{x}^{(1)})\left[f_1(\mathbf{x}^{(j)})-f_2(\mathbf{x}^{(j)})\right]<0.
\end{equation}
This equation is equivalent to the ISO condition for $2$-period orbits.

Thus, it appears that a $2$-period orbit is ISO irrespective of its stability property (and hence Proposition 1 naturally follows from Proposition 2). However, the fraction of mutants required to invade ISO has to be comparatively larger for the case of stable $2$-period orbits. It must be kept in mind that the analytically tractable case of the $2$-period orbits of the replicator map appears to be very special in this respect: It remains to be proven that how much of this discussion goes over to the higher period orbits in arbitrary escort-incentive dynamics.

\bibliography{Bhattacharjee_etal_bibliography}
\end{document}